\documentclass[12pt,preprint]{aastex}

\slugcomment{\bf to appear in The Astrophysical Journal}

\begin{document}

\title{Kinematics of Ultracompact HII Regions Revealed: \\ High Spatial
and Spectral Resolution Mapping of the 12.8 $\mu$m [Ne~II] Line in Monoceros R2}

\author {D.T. Jaffe 
\altaffilmark{1} 
\altaffilmark{2}
, Q. Zhu  
\altaffilmark{1}
, J.H. Lacy 
\altaffilmark{1}
, M. Richter 
\altaffilmark{1}
\altaffilmark{3}
}
\affil{Department of Astronomy}
\affil{University of Texas at Austin, Austin, TX 78712}
\altaffiltext{1}{Visiting astronomer at the Infrared Telescope Facility, 
which is operated by the University of Hawaii under contract from the
National Aeronautics and Space Administration.}
\altaffiltext{2}{Guest Scientist, Max-Planck-Institut f\"ur extraterrestrische
Physik, Garching, Germany.}
\altaffiltext{3}{Department of Physics, University of California, Davis}


\begin{abstract}

We present the first results of a study of the kinematics and
morphology of ultracompact H~II regions using a new observational
technique.  We used very high spatial and spectral resolution
observations ($\theta_{Beam}$=1.4\arcsec, $\Delta$V= 3.4 km s$^{-1}$) of
the [Ne~II] line at 12.8 $\mu$m to study the ionized gas in 
Monoceros R2.
The [Ne~II] emission shows an H~II region
with highest emission measure in a $\sim$24 \arcsec \ diameter
shell,
surrounded on all sides by neutral material.
[Ne II] line widths are as narrow as 8 km
s$^{-1}$ at some positions. In 
places where the lines are complex and broader,
the additional width is most likely due to overlap
of narrower features along the line of sight. 
The
narrow features themselves, however, are broader than
the purely thermal width. 
The global kinematics suggest that the 24\arcsec \ shell is
expanding at $\approx$10 km s$^{-1}$.  
This interpretation leads 
to a dynamical age for the H~II region of $<$10$^4$ years.
However, the spectral profiles toward the brightest part of the nebula
(on the southeast side of the shell) are not consistent with a simple 
expansion picture.
Both the 24\arcsec \ shell and the bright southeast ridge can be part of a
common kinematic pattern in which material flows from the bottom
to the rim of a bowl-like feature.

High resolution
observations of mid-IR fine structure transitions offer great promise
as a probe of the kinematics and morphology of
ionized regions around very young massive stars.
Once appropriate theoretical modelling and observations of a larger
sample of UCHII regions are in place, it should be possible to
determine the physics behind the observed systematic
motions in sources like Mon R2.

\end{abstract}





%

\section{Introduction}

This study presents the first results of a program to examine ultracompact
H~II regions at high spatial and spectral resolution using fine-structure
line emission in the mid-infrared.  This kind
of observation allows us to see more clearly the kinematics and structure
of these ionized regions around very young, massive stars. We illustrate the technique
by investigating the properties of the H~II region in the core of the
Monoceros R2 molecular cloud.

\subsection{Ultracompact H~II Regions}

Massive stars evolve very rapidly to the zero age main sequence.  They begin
to emit copious amounts of Lyman continuum radiation while still embedded
within their natal clouds.  When this ultraviolet emission turns on, a dense
ionized region forms.  These ultracompact 
H~II regions (UCHII regions) are one of the earliest
manifestations of newly formed massive stars.
High frequency radio interferometric
surveys, together with studies of IRAS point sources with characteristic
colors, show that there are several hundred UCHII regions in our
galaxy (Wood \& Churchwell 1989a, Garay et al. 1993, Kurtz et al. 1994).  
These sources range in size down to 0.005 pc.

High spatial resolution radio continuum studies reveal a range
of morphologies for UCHII regions (Wood \& Churchwell 1989b).  
Almost half of such sources are
spherical or unresolved at 1-2 arc second resolution with the VLA.
The other half of the sources are almost evenly divided
between cometary shapes, core-halo sources, and irregular sources, with
a small percentage (about 4\%) having shell morphologies.  
The morphological
classifications of many sources, however, need to be viewed with caution.
Not only does the limited signal-to-noise of the VLA ``snapshot''
surveys keep observers from seeing lower surface brightness components
of the target sources, but
undersampling of the spatial frequency domain by aperture
synthesis telescopes also can affect the appearance of the sources
(Wood \& Churchwell 1989b, Fey et al. 1992). 
In addition, the UCHII regions often lie
in regions with significant
amounts of continuum emission over a large area.  The interaction of
this extended complex emission with the uv sampling of the VLA can
limit the dynamic range of the radio continuum
observations and make it hard to draw conclusions with confidence
from any but the brightest portions of the maps.

The high densities in UCHII regions lead to enormous 
(nT$\sim$10$^9$ cm$^{-3}$ K) internal
pressures. If the smallest sources expand at their internal
thermal sound speed into the surrounding interstellar medium, their
expansion timescales can be as short as a few hundred years (Dreher \&
Welch 1981).  The number
of UCHII regions and the short expansion timescales combine to give
rise to what is known as the ``UCHII region lifetime problem'':  Extrapolation
from the number of UCHII regions and the thermal expansion timescales
of the individual regions leads to excessively high star formation rates
for massive stars (cf. Wood \& Churchwell 1989a).  
A number of solutions to the lifetime problem
have been proposed for various morphological types of UCHII regions.
These include: a model for cometary UCHII regions employing
bowshock compression by O stars moving relative to the
neutral cloud (Van Buren \& MacLow 1992), 
explanations of unresolved and bipolar
regions as mass-loaded wind sources (Redman, Williams \& Dyson 1995, 1996,
Dyson, Williams, \& Redman 1995, Lizano et al. 1996) or as photoevaporated
circumstellar disks (Hollenbach et al. 1994, Yorke \& Welz 1993, 1996, 
Richling \& Yorke 1997, Jaffe \& Martin-Pintado 1999),
and the suggestion that extremely dense and warm neutral material may
pressure confine the UCHII regions (De Pree, Rodriguez, \& Goss 1995a,
Akeson \& Carlstrom 1996).  
Each of these ideas has
limited applicability and each has problems, even for the source types
where it works best.  The date we present in this work lead to new
ideas about the lifetime problem in Mon R2.

Our ability to test the models of UCHII regions depends on good information
about the morphology and, even more importantly, the kinematics of the
sources. 
The kinematic information we currently have about
UCHII regions comes from observations of centimeter-wave hydrogen
recombination lines.  The nature of the emission, however, limits the
diagnostic power of the lines.
The thermal widths of the hydrogen lines make it hard to detect
bulk motions of the gas unless these motions are a significant
fraction of the thermal sound speed.
Measurements of
lines arising from states with principal quantum number n$>$100
may also suffer from pressure broadening making the higher frequency,
lower n lines better tracers of kinematics.
Despite these problems, the hydrogen recombination line emission shows
evidence for gas motions both
on the scale of the entire region, through shifts in velocity centroids
across the source in some sources
(Garay et al. 1986, De Pree et al. 1994, 1995b, Garay et al.
1994), and on a scale smaller than the beamsize, which
can be as small as 0.2\arcsec-1\arcsec, through the large local 
line widths seen in most UCHII regions.
The line widths seen in individual
beams (25 to as much as 180 km s$^{-1}$, Garay \& Lizano 1999)
are usually attributed to combination
of thermal motion, turbulent motion, and unresolved mass motions.
If the thermal and other components can be added in quadrature, 
the large observed
line widths imply that the turbulence and/or mass motions alone
would produce line widths $>$20 km s$^{-1}$. 

The VLA observations of hydrogen recombination lines suffer from 
the same uv sampling problems as the continuum studies.  In addition,
they usually have fairly coarse velocity resolution (typically 10 km
s$^{-1}$, although the VLA backend is capable of finer resolution)
and sensitivity-limited dynamic range of only 20 or so in a given
velocity channel. The radio recombination line measurements
therefore tend to miss both the larger-scale emission within the individual
regions and also small-scale features if they are weak compared
to the brightest spectral features in the region.  

\subsection{High Resolution Mid-IR Fine-Structure Line Mapping}

The $^2$P$_{1/2}\rightarrow^2$P$_{3/2}$ transition of Ne$^+$ is a
versatile tracer of the ionized gas in H~II regions.  At a frequency
of 780.4238 cm$^{-1}$ (12.81 $\mu$m, Yamada, Kanamori, \& Hirota 1985)
this line lies in a part of the mid-infrared spectrum where both interstellar
dust and the Earth's atmosphere are relatively transparent.  
Since Ne$^{\rm o}$ and Ne$^+$ have ionization potentials
of 21 and 41~eV, Ne$^+$ is the dominant neon ion in H~II regions with a broad
range of stellar types for their exciting stars.  The critical density
of the 12.8 $\mu$m [Ne~II] transition is 5.4$\times$10$^5$ cm$^{-3}$,
so the line intensity scales with the emission measure in all but the
densest parts of the H~II regions. Earlier observations have shown
that the [Ne~II] distribution closely resembles the distribution of
short wavelength radio continuum emission (Lacy, Beck, \& Geballe 1982,
Beck, Lacy \& Kelly 1998, Takahashi et al. 2000).

Unlike the hydrogen line observations, the observations of [Ne~II] 
give us the ability to see subsonic bulk
motions of the gas. In a nebula with a kinetic temperature of 10$^4$ K,
the recombination lines of hydrogen have a purely thermal line width
of 21 km s$^{-1}$, full width to half maximum (FWHM, see Gordon 1988).
If the gas is otherwise at rest, the atomic mass of neon (20) means
that the 12.8 $\mu$m line will be a factor of 
(m$_{\rm Ne}$/m$_{\rm H}$)$^{1/2}$= 4.5 narrower than the hydrogen
lines, or 4.7 km s$^{-1}$ at 10$^4$ K.
The significantly narrower thermal line width
of the neon profiles means that high resolution spectroscopy of
the 12.8~$\mu$m line can offer us the chance to resolve 
closely-spaced velocity components along a given line of sight,
to follow small variations in line velocity from one position to
another, and to detect the presence of turbulence or bulk motion
on small scales but with overall velocity spreads significantly
below the thermal sound speed.

\subsection{The Ultracompact HII Region in Monoceros R2: A Case Study}

The Monoceros R2 complex is a nearby (D=830 pc, Herbst \&
Racine 1976) region currently forming high and low mass stars.  The core
of the complex contains an ultracompact H~II region, $\sim$0.1 pc
(24\arcsec) in diameter
with a bright ridge along its southeastern side, as seen in
radio continuum emission (Massi, Felli, \& Churchwell 1985, Wood \& 
Churchwell 1989b) and Br$\alpha$ (Howard et al. 1994).  Although there is
additional emission from a more diffuse ionized component (Downes et al.
1975), the majority of the flux arises in this UCHII region. Downes et
al. conclude that a star of ZAMS spectral type B0 or earlier is needed
to provide sufficient ionizing radiation to the HII region. Massi
et al. interpret the morphology of the ionized gas emission as a blister
HII region illuminated by a source close to the bright arc.

The inner arcminute of Mon R2 also contains a group of bright
near-IR point sources (Beckwith et al. 1976, Aspin \& Walther 1990,
Howard et al. 1994).
The total luminosity emitted within this region is around 
5$\times$10$^4$ L$_{\odot}$ 
(Thronson et al. 1980).            
Near-infrared imaging of the Mon R2 region on 15\arcmin \ scales 
reveals that the nebula
lies within the core of a cluster containing at least 400 stars
and that the stellar density peak is roughly coincident with IRS2,
the bright source inside the northern part of the nebula (Carpenter
et al. 1997).            

High spatial resolution (10\arcsec--14\arcsec) mapping of molecular 
line emission
from the Mon R2 shows that dense (up to 
n$_{\rm H2}$=3$\times$10$^6$ cm$^{-3}$, Choi et al. 2000) 
gas surrounds
the UCHII region on three sides
(Torrelles et al. 1990, Gonatas, Palmer, \& Novak 1992,
Tafalla et al. 1994, Giannakopoulou et al. 1997). The
highest resolution observations (Gonatas et al. 1992) show a ring of
molecular emission immediately outside of the region of bright radio
continuum emission.  
On larger scales, observations of $^{12}$CO rotational lines show that
the UCHII region in Monoceros R2 sits near the middle of a huge
(several parsec long) and massive (a few 10$^2$ M$_{\odot}$) bipolar
outflow (Wolf, Lada, \& Bally 1990,  Meyers-Rice \& Lada 1991).

This paper presents and analyzes
a detailed [Ne~II]
map of a 45\arcsec \ by 28\arcsec \  region containing 
the bright radio continuum
source in Mon R2 and hopes to demonstrate the value of high resolution
observations of UCHII regions using fine-structure lines in the mid-IR.  
In Section 2, we discuss the instrumentation,
observing technique and
data analysis necessary to produce high signal to noise, high spatial and
spectral resolution observations of the [Ne~II] line. In Section 3, we
present the results of the [Ne~II] spectral line mapping and describe the 
major features of the resulting datacube.  Section 4 interprets the 
kinematics of the ionized gas and draws implications for the nature and
structure of the nebula.

\section{Observations}

\subsection{Instrument and Observing Technique}

We mapped the [Ne~II]  line
toward the ultracompact
H~II region Mon R2 in 2001 November with the Texas Echelon Cross Echelle 
Spectrograph (TEXES, Lacy et al. 2002) on the 3m NASA Infrared Telescope
Facility on Mauna Kea.  

In the setup used for these observations, the slit was oriented 
north-south, had a width of 1.4\arcsec \
and a length of 11.5\arcsec.  With the chosen slit width, the instrument
produces a cross-dispersed spectrum with a resolving power of 88,000
(velocity resolution of 3.4 km~s$^{-1}$) at 12.8 $\mu$m.
The pixel spacing of the 256$^2$ Si:As array
was 0.35\arcsec \ on the sky (0.95 km~s$^{-1}$ in the spectral direction).  
The diffraction limit
of the IRTF at 12.8 $\mu$m is 0.88\arcsec.

We mapped Mon R2 by stepping the telescope from west to
east in 0.7\arcsec \ increments along 45\arcsec \ long strips across
the nebula.  The integration time at each position along each strip was
1 second. We observed 5 west-east strips separated by 5 arcseconds
north-south.  Since each strip was scanned four times, the integration
time per point in the map was approximately 9 seconds.
The spectral and spatial variations in the
sky background are normally small enough when 
observing at high spectral resolution
that rapid subtraction of emission at a reference position is not necessary.
For the Mon R2 maps, we therefore held the telescope secondary fixed
and
were able to use spectra taken at the ends of each scan, at positions
where no emission was present, as sky references.

We made two independent determinations of the absolute
position of our [Ne~II] map.  In the first method, we matched the position
of the bright southeastern ridge and the point source near IRS1 to the
corresponding features in the radio continuum map of Massi et al. (1985).
Our second method matched the position of the 12.8 $\mu$m continuum source 
IRS2 with the near-IR position of the source (Aspin and Walther 1990).
The two methods agree to better than 2\arcsec \ in R.A. and 0.\arcsec7 in 
Declination.
The average of the two solutions places the (0,0) point of our map at 
06$^h$07$^m$45.9$^s$, -06$^{\rm o}$23\arcmin 01\arcsec \ (J2000).

\subsection{Data Reduction and Analysis}

The spectra were reduced using the TEXES automated pipeline reduction
program which removes artifacts, flat-fields the frames, and extracts
the spectrum from the cross-dispersed data (Lacy et al. 2002).
We apply a radiometric flux calibration using measurements of
ambient temperature and cold loads before each set of scans.
The intensity scale is in units of erg cm$^{-2}$ s$^{-1}$ sr$^{-1}$
(cm$^{-1}$)$^{-1}$.  The wavelength solution was derived from
telluric lines present in each spectrum and is accurate to
$\sim$1 km s$^{-1}$.

\section{Results}

\begin{figure}
\epsscale{0.65}
\plotone{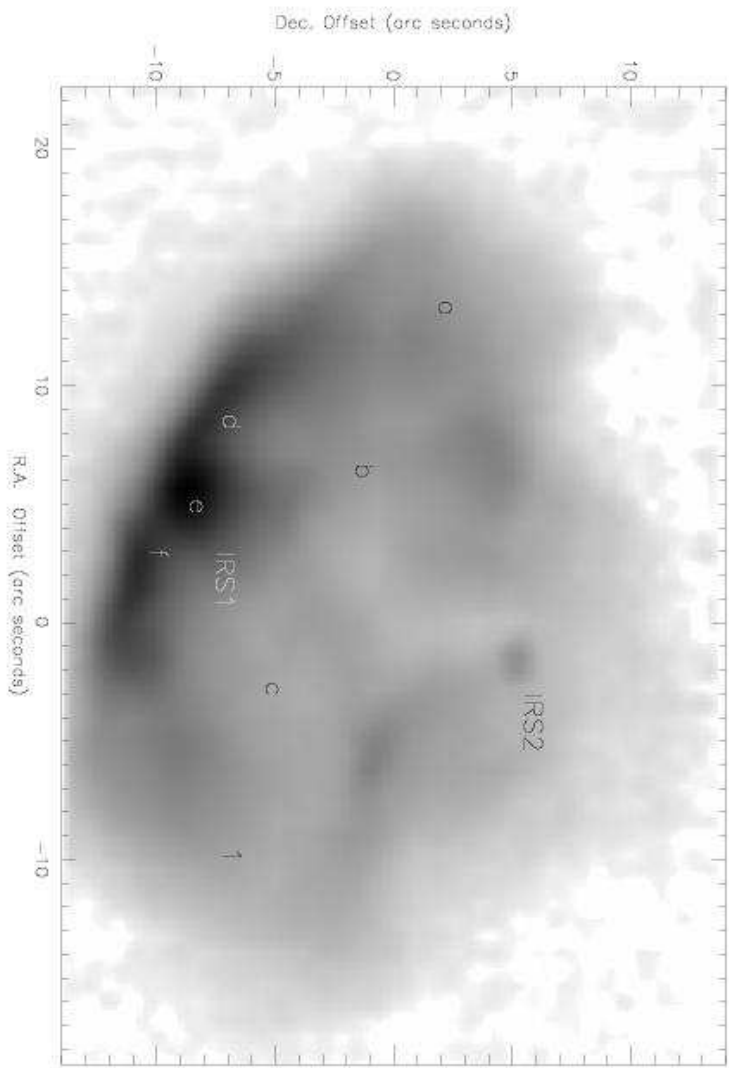}
\caption[1]
{\label{integ} Integrated line brightness distribution for the [Ne~II]
line toward Monoceros R2.  The (0,0) position for this and all
other maps is
06$^h$07$^m$45.9$^s$, -06$^{\rm o}$23\arcmin 01\arcsec \ (J2000).
The map
includes the sum of channels between -7.3 km s$^{-1}$ and 35.3 km s$^{-1}$.
The greyscale displays the square root of the intensity in the [Ne~II]
line with white corresponding to 0 and black to 0.28 erg cm$^{-2}$ sec$^{-1}$
sr$^{-1}$.
 The 
point source at (-2,$+5$) is produced by 12.8 $\mu$m continuum
emission from IRS 2 whereas the bright source at ($+5.5$,-8) is
largely [Ne~II] line emission from the region near IRS1. The number '1'
shows the position of the spectrum shown in Figure \ref{narrow}. 
The letters 'a' to 'f' give the positions of the spectra in 
Figure \ref{spectra}.
}  
\end{figure}

Figure \ref{integ} shows a map of [Ne~II] integrated line brightness 
for the Mon R2
region. The neon emission extends over $\approx$ 24\arcsec $\times$ 18\arcsec. A
bright ridge or flattened arc
in the southeastern part of the source, running from southwest
to northeast, dominates the emission.  This ridge lies about 4\arcsec \
to the southeast of the near-IR continuum source IRS1.  At lower flux
levels, the ridge extends to form a complete shell surrounding a region
of lower brightness.  The faintest part of the shell in [Ne~II]
lies east of IRS2, 
but there is detectable emission out to 
$\sim$10\arcsec \ from the center of the nebula in all directions.
The flux distribution is quite symmetric about an axis 
at position angle -25$^{\rm o}$ and running almost 
through the positions of
Mon R2 IRS1 in the southeast and IRS2 in the north.
The overall morphology
agrees very closely with the radio continuum flux distribution (Massi et al.
1985, Wood \& Churchwell 1989b). The more limited [Ne~II] imaging of
Takahashi et al. (2000) is consistent with the map shown in
Figure \ref{integ} . 

\begin{figure}
\epsscale{1.0}
\plotone{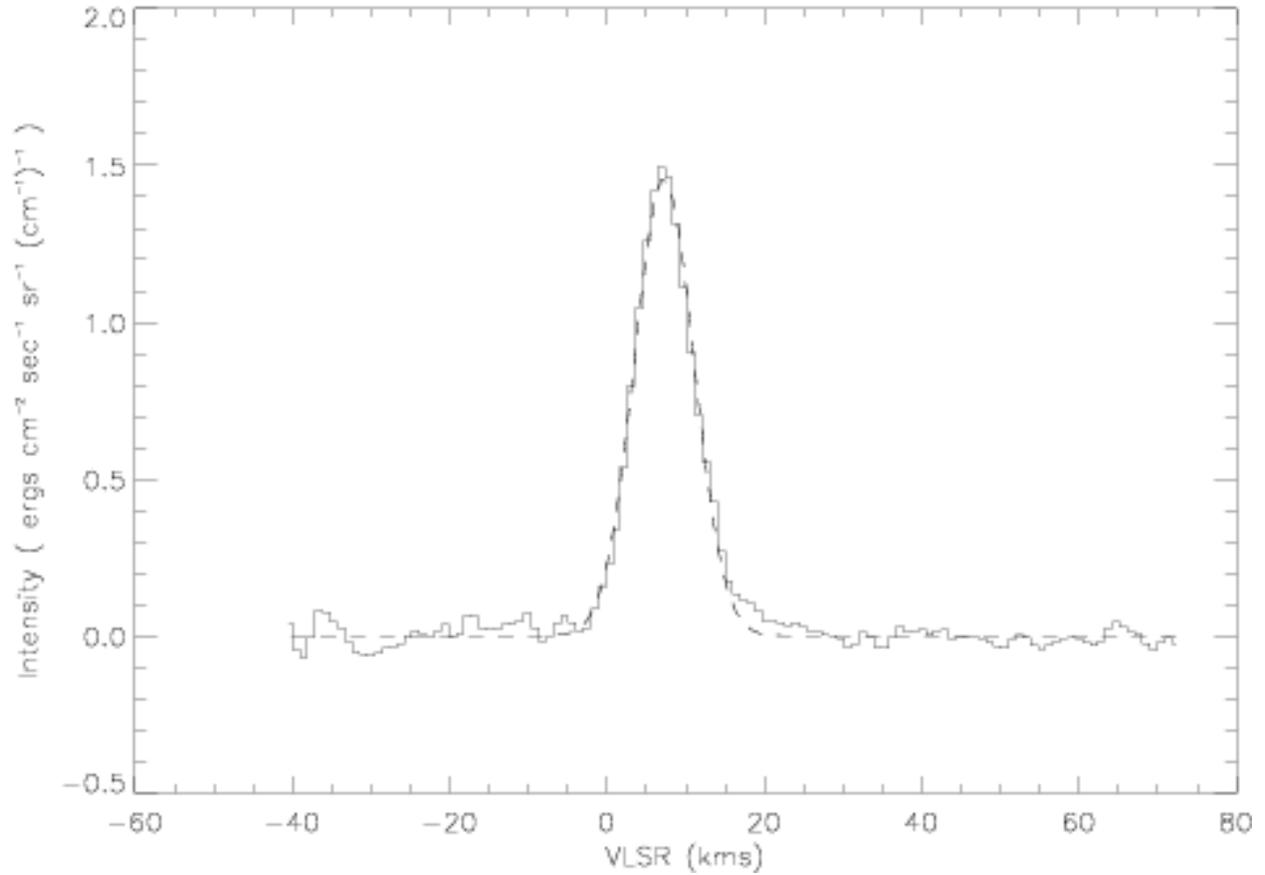}
\caption[3]
{\label{narrow} [Ne II] spectrum at a position
along the edge of the nebula in the southwest
(-9.5\arcsec, -7.2\arcsec \  from the map center) (solid line). 
A Gaussian fit to this
spectrum (dashed line) shows that the full width to half maximum of the
line at this
position is only 8.8 km s$^{-1}$. The position where this spectrum
was taken is labeled as '1' in Figure \ref{integ}.
}
\end{figure}

\begin{figure}
\epsscale{1.0}
\plotone{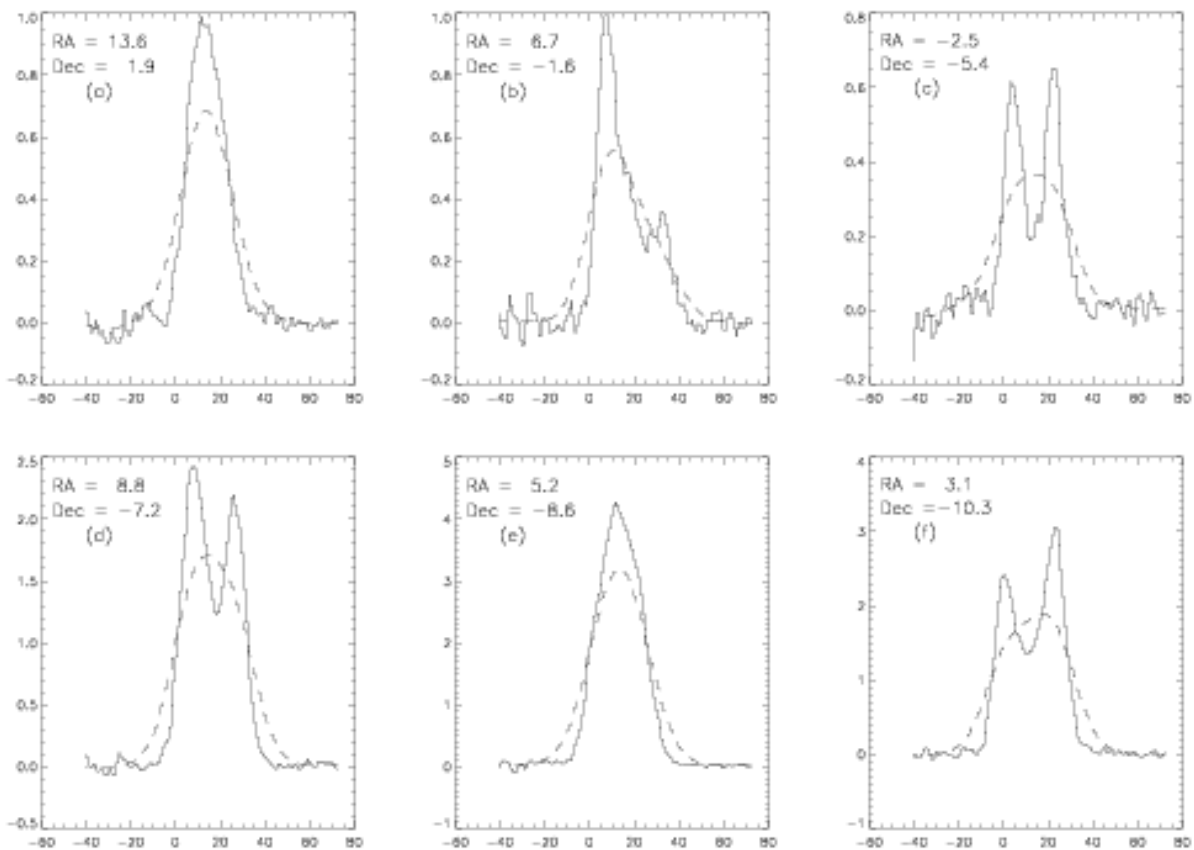}
\caption[4]
{\label{spectra} Sample [Ne~II] spectra from different parts of the
Mon R2 nebula.  The solid line shows the observed [Ne~II] spectrum
at full ($\Delta$V= 3.4 km s$^{-1}$) resolution.  The dashed lines
show the same spectra smoothed with a 20 km s$^{-1}$ Gaussian
(the thermal broadening of hydrogen lines at 10$^4$ K).
The inset coordinate gives the offset in arc seconds from
the (0,0) position.  The top row shows spectra across the nebula from
the northeast to southwest (labeled as 'a-c' in Figure \ref{integ}). 
The bottom row contains spectra from the bright southeastern ridge
(labeled as 'd-f' in Figure \ref{integ}).
The x axis displays V$_{\rm LSR}$ in km~sec$^{-1}$
while the Y axis in each figure is intensity in erg cm$^{-2}$ sec$^{-1}$
(cm$^{-1}$)$^{-1}$.
}
\end{figure}

The sensitivity of TEXES is good enough to allow us to explore the 
details of the kinematics in Mon R2.  For example, TEXES observed 
the S/N$\sim$50 
spectrum shown in Figure \ref{narrow}, in $\approx$9
seconds of integration time.  This spectrum was taken well away
from the brightest parts of the nebula, 
in a region where the 6~cm continuum flux in a 1.4\arcsec \ beam
is $\sim$10 mJy (Massi et al. 1985).
 At the edges of 
the nebula away from the southeastern
ridge, the lines are typically single and have 
FWHM$\sim$9-14 km s$^{-1}$. The spectrum shown in 
Figure \ref{narrow} was taken at a position along
the southwest side of the nebula (labeled as '1' in Figure \ref{integ})
 while the upper left panel of Figure
\ref{spectra} shows a spectrum from the northeast rim (labeled 'a' in
Figure \ref{integ}).
Along most of
the bright southeastern ridge and throughout the region inside
the shell, the spectra are double-peaked. The two remaining panels in 
the upper row of Figure \ref{spectra} show profiles from the interior
of the nebula ('b' and 'c' in Figure \ref{spectra}) 
while the left and right panels in the lower row
show profiles $\sim \pm$3\arcsec \ to the left and right of the center
of the bright southeastern ridge ('d' and 'f'). 
Near IRS1 ('e'), at some
positions along the bright ridge, the two components merge to form
a broader (27 km s$^{-1}$ FWHM) line with a flattened 
top.  At the lowest detectable levels, the [Ne~II] lines along the
southeast ridge and in the center of the nebula span a range of
40--50 km s$^{-1}$ with no evidence for higher velocity emission
at lower levels.

\begin{figure}
\epsscale{1.0}
\plotone{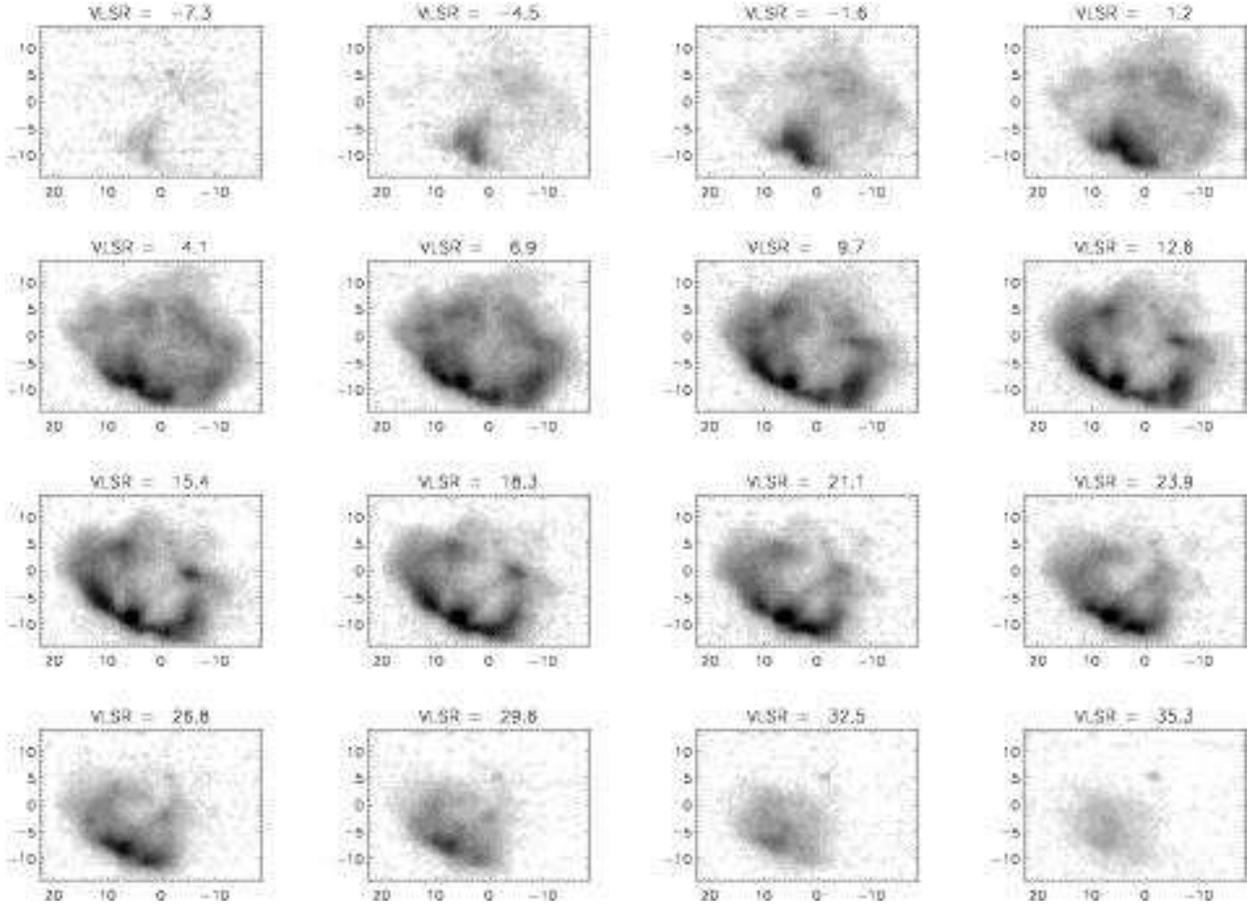}
\caption[2]
{\label{chan} Distribution of [Ne~II] emission toward Monoceros 
R2 at different radial velocities.  Each subfigure is the sum of
three spectral channels from our data cube. The spectral response
function causes a slight overlap between the bins.
The intensity scale runs from white = 0 to black = 3 erg cm$^{-2}$
sec$^{-1}$ sr$^{-1}$ (cm$^{-1}$)$^{-1}$ in all panels.
}  
\end{figure}

We need to visualize the 
extensive dynamical information made available by the [Ne~II]
data cube in a variety of ways.  Figures 
\ref{chan} -- \ref{pvlr} 
 present channel maps at selected velocities and
position-velocity (PV) diagrams along cuts through the nebula. 
For the PV diagrams in Figures \ref{pvud} -- \ref{pvlr}, 
we have rotated the spatial plane in the data
cube by -25$^{\rm o}$ so that the symmetry axis of the nebula is
vertical. The PV cuts in Figures \ref{pvud} 
and \ref{pvud2} run parallel to the axis of symmetry
and those in Figure \ref{pvlr} perpendicular to
this axis.

The channel maps and PV diagrams show that the phase space
within our data cube is not filled randomly.  The channel maps show
a small source in the southeast at the extreme velocities (-7 km s$^{-1}$
 to -4.5 km s$^{-1}$ and 32  km s$^{-1}$ to 35  km s$^{-1}$), 
emission in the nebular interior at slightly lower velocity offsets,
and a growing-then-shrinking
ring and well-developed central hole at intermediate
velocities.  The `vertical' PV cuts in Figures \ref{pvud} and \ref{pvud2}
show single lines along the northeast and southwest borders of the nebula.
Cuts through the center of the nebula show that,
inside the bright ring, there are usually two intensity 
maxima at velocities separated by about 20  km s$^{-1}$.
There is some indication in the PV cut through the center of the nebula
(upper panel of Figure \ref{pvud2}) for a velocity gradient from northeast 
to southwest across the
nebula of 5-7  km s$^{-1}$.

The PV cut along the bright southeast ridge
(left-hand panels of Figure \ref{pvlr})  expands on
the picture provided by the spectra in the lower row of Figure \ref{spectra}.
An almost perfect ring centered at ($\Delta$RA=--2\arcsec, 
V$_{LSR}$=17 km s$^{-1}$)
with a diameter of (12\arcsec, 27  km s$^{-1}$) is crossed by a $\sim$4\arcsec \
wide bar of broad-line emission that spans the entire velocity range
of the ring. The bar is canted in the spatial direction 
by about 2\arcsec \
between its two extremes in velocity.

\section{Discussion}

Our spectral mapping of the 12.8 $\mu$m [Ne~II] line in Monoceros R2
demonstrates the unique ability of such high spatial and spectral resolution 
observations to 
trace both the distribution and kinematics of the ionized material
in ultracompact HII regions. 
The close correspondence of the
[Ne~II] and radio continuum maps argues that the neon line faithfully 
traces the emission measure.  The channel maps and spectra show
that the 12.8 $\mu$m line can also
disentangle the distribution of emission measure among various kinematic
components along a given line of sight.

The interstellar extinction toward Mon R2 at 12.8 $\mu$m is modest.  Natta
et al. (1986) use the Brackett decrement to derive an extinction toward
the nebula of A$_{4.05 \mu m}$=1.1  
(A$_{\rm V}\approx$33, A$_{12.8 \mu m}\approx$0.7). 
Comparisons of Br$\alpha$ and radio continuum maps
and of Br$\alpha$ and Br$\gamma$ maps show that the highest extinction 
lies along a north-south ridge from north of the southeast ridge to
a region beyond the northern edge of the nebula (Howard et al. 1994).
With the exception of this central ridge, extinction will not 
prevent the [Ne~II] line strength
distribution from accurately tracing the ionized
material.

\subsection{Local Kinematics}

The neon lines at some locations in Mon R2 are extremely narrow.
At the position of the spectrum in Figure \ref{narrow}
(9.5\arcsec \ W, 7\arcsec \ S of the map center),
the FWHM of the neon line is 8.8 km s$^{-1}$. Deconvolving this
line with our spectral resolution (3.4 km s$^{-1}$), 
the FWHM is only 8 km s$^{-1}$ .
This measured [Ne~II] line width is 0.4
times the width of a thermally broadened hydrogen line in 10$^4$ K
gas though larger than the thermal width of a neon line in the same
gas (4.7 km s$^{-1}$).
At neighboring positions, the line is almost as narrow.
This width is smaller than the line widths measured toward any
UCHII region using high spatial and spectral resolution observations
of hydrogen recombination lines.  
 
Where the neon lines are broader
(Figure \ref{spectra}), the lines clearly consist of multiple
narrow peaks with typical widths for the individual features of
10--14 km s$^{-1}$.  
At those few locations with broad single lines, the non-Gaussian
line shapes and the kinematics of the surrounding regions argue
that the line shapes result from the overlap of narrower
velocity  components along the line of sight.
At FWHM=10--14 km s$^{-1}$, the
individual features in the [Ne~II] spectra are still significantly
wider than the thermal width for this line in 10$^4$ K gas.  There
are a number of possible contributions to this additional line width:
In some parts of the nebula, velocity gradients across the beam add
up to 2 km s$^{-1}$ to the line width (see the bottom right panel 
of Figure \ref{pvlr}). If the large-scale features at $\sim$7 km
s$^{-1}$ and $\sim$27 km s$^{-1}$ each consist of ionized material associated
with several large-scale surfaces with different velocities, the
blending of features will broaden the lines.  Neutral gas with spatially
extended emission from two velocity components separated by a 
few km s$^{-1}$ is common in giant molecular clouds.  If the neutral
material is at a single velocity, the lines from ionized gas can be broadened
beyond their thermal widths if the ionized material streams off of clumps
in several directions.  In addition, it is possible that turbulence in the 
ionized gas produces the 4-10 km s$^{-1}$ of broadening seen in the
[Ne~II] lines. 
 
The conclusion that the local line widths throughout the Mon R2
UCHII region are narrow differs from the conclusions drawn
from studies of radio
recombination lines from compact H~II regions since the earliest days
(Mezger and Hoglund 1967) and
from those drawn from most radio
recombination line studies of UCHII regions
observed  over
the past two decades (see Afflerbach et al. 1996, Garay \& Lizano 1999).
These earlier hydrogen line studies typically imply contributions to observed
line widths by local
turbulence of 20--30 km s$^{-1}$.
In the hydrogen recombination lines, blends of multiple components 
separated by 5--10 km s$^{-1}$ will look much more like simple Gaussian
profiles.  The difference between thermal widths of 4.5 km s$^{-1}$ for
neon and 20 km s$^{-1}$ for hydrogen
therefore completely alter the way we perceive the 
physical cause of the observed line profiles.
The spectra shown as dashed lines in Figure \ref{spectra} illustrate
this point.  At each position, we have smoothed the observed [Ne~II]
spectrum with a 20 km s$^{-1}$ FWHM Gaussian to simulate the profile
we would observe in a themally-broadened hydrogen line arising in 
10$^4$ K gas.
These smoothed profiles appear quite Gaussian to the eye, even when
the underlying [Ne~II] profile is distinctly double-peaked with a 
substantial velocity separation between the components.                                               
\subsection{Global Morphology, Expansion and Dynamical Lifetime of the Nebula}

At high intensity levels, Mon R2 has the morphology of a 
blister, broken shell, or cometary nebula (it was classified as cometary in
the radio continuum survey of Wood \& Churchwell 1989b).
Our [Ne~II] results are consistent with earlier radio continuum and
10 $\mu$m continuum mapping in showing strongly limb-brightened
emission
on three sides, all except the northwest (Figure \ref{integ} of this paper,
Massi et al. 1985, Telesco 2002 (personal communication, available as:
\\ http://www.astro.ufl.edu/oscir/monr2\_n.gif)).
In the H and K band, where polarization studies show that most of the
 continuum is scattered light
(Aspin \& Walther 1990, Yao et al. 1997),
the source is limb-brightened
at all azimuthal angles (Aspin \& Walther 1990).
The polarization mapping demonstrates a direct connection between
the ionized and neutral gas.  The circular symmetry of the polarization
vectors over most of the nebula imply that light from IRS2 
scatters off the walls of a hollow cavity and into our line of sight.
The lower degree of polarization in the southeast argues that the 
local source IRS1 contributes a substantial fraction of the scattered
light along the southeast ridge.  
The low
point in the near-IR shell intensity just to the east of IRS2
probably results from obscuration 
since it lies along the
lane where Howard et al. (1994) had inferred significant near-IR
dust extinction. 
In the 3.29 $\mu$m PAH feature (Howard et al. 1994), 
there is a complete ring of emission 
outside of the ionized nebula, implying that there is neutral material
on all sides of the H~II region, in agreement with the millimeter
HCO$^+$ mapping results (Gonatas et al. 1992). 
The uniformity of this ring indicates
that the obscuring dust lane does not extend to larger radii.

\subsubsection{Spherical Expansion and Nebular Lifetime}

The neon emission at velocities near the mean nebular velocity
(Figure \ref{chan}) forms a well-defined ring.             
Along and within the ring, the neon lines show a 
pattern of radial motion with an expansion or outflow velocity of 
$\approx$ 10 km s$^{-1}$. 
The pattern in the channel maps where the center of the source
dominates at the extreme velocities and a ring forms at intermediate
velocities, as well as the position-velocity
diagrams in Figures \ref{pvud} -- \ref{pvlr} supports this conclusion.

We can use the measured size and expansion velocity of Mon R2 to estimate the
dynamical age of the HII region.  
A naive crossing time argument implies an age
of 4$\times$10$^3$ years. This time shows that Mon R2 shares the 
problem of the UCHII region population as a whole in having a dynamical
lifetime that would predict an OB star formation rate 1-2 orders of
magnitude higher than the rate inferred from counts of the stars 
themselves. This lifetime
is also significantly shorter than the age one estimates
using the formalism of Spitzer (1978) which calculates the size of the
H~II region as a function of time assuming uniform expansion into
a static neutral cloud after
the initial ionization has reached equilibrium.
Under this assumption, an HII region excited by a B0V star 
producing
10$^{47.8}$ Lyman continuum photons per second (Downes et al. 1975)
and having a surrounding
neutral medium with a density of 10$^6$ cm$^{-3}$ would reach the
measured size of Mon R2 in 1.8$\times$10$^4$ years.
This higher value is closer to but still less than the average value 
needed for consistency with the OB star formation rate.

\subsubsection{Refining the Kinematic Model}

Examined more closely, not all of the 20\arcsec \ diameter ionized shell 
displays the simple velocity pattern of a radially expanding
spherical shell.  The two most notable differences
are the lack of curvature in the vertical position-velocity cuts
through the center of the nebula (Fig. \ref{pvud} (bottom) and
\ref{pvud2} (top)) and the ``wrong way sign''
velocity pattern along the bright
southeastern ridge (Fig. \ref{pvlr}).
 
Cuts through a symmetrically expanding spherical shell should
show a maximum velocity splitting in
the middle of the nebula and should gradually approach the cloud velocity
as one reaches the rim.
The vertical cuts in the bottom panel of Figure \ref{pvud} and the top
panel of Figure \ref{pvud2} indicate that the velocities of the two
components have very little curvature from south to north through the inner
part of the nebula and instead display a straight-line connection 
between two
broad-line regions at the nebular rim. The velocity difference between
the two components tends to decrease, however, going north from the bright
southeast ridge (Fig \ref{pvud2} (top)).
 
A horizontal cut through the bright southeastern ridge
(Fig. \ref{pvlr}) differs even more from the simple expansion model.
Instead of the narrow line at the cloud velocity that we see in the
vertical cuts along the northeast and southwest sides of the nebula
and which we would expect in a spherical expansion,
this cut shows a ``wrong-way sign''; a 
ring in PV space bisected by a broad spectral
feature.  The broad feature is bipolar, with a barely resolved
($\approx$2\arcsec) tilt and a velocity extent
of 20 km s$^{-1}$. The core of this feature appears in the
channel maps (Figure \ref{chan}) as
a bright knot $\approx$2\arcsec \ southeast of IRS1 present from
$\sim$7--24 km s$^{-1}$.     
 Cuts within a few arc seconds north and south of the one
shown in Figure \ref{pvlr} exhibit similar behavior.

The velocity pattern shown in the position-velocity cuts in Figures
\ref{pvud} -- \ref{pvlr} fits fairly well to a simple kinematic
model in which the gas flows along the surface of a bowl-shaped
nebula with the flow moving from the base of the bowl
(just below IRS1) upward toward the rim. 
The flow originates near the southern 
end of the nebular symmetry axis, to the southeast of IRS 1,
and flows out along the bright southeastern ridge.  The circular
PV feature in the horizontal cut in Figure \ref{pvlr}
shows the initial outward motion of the gas.
As the material flows upward along the curved walls of the bowl, 
the vertical cuts in Figures \ref{pvud} and \ref{pvud2} display
split lines. 
Once these walls become more or less vertical,
the velocities of the features should drop toward the cloud velocity.


The broad-line feature at the center of the P-V diagram in Figure
\ref{pvlr} either represents unresolved expansion at the apex
of the flow or a separate feature due to activity unique to the
dense gas immediately adjacent to IRS1.
Although this source appears to be $\approx$2\arcsec \ southeast of
IRS1, the positional uncertainties allow for the possibility that 
the two objects coincide.  If the broad line region does coincide
with IRS1, it may represent the ionized
portion of a less collimated wind that only has a high density
close to the  exciting star of the
outflow. If this flow is fully ionized and moves at constant
velocity so that density is proportional to r$^{-2}$, the
projected intensity in the [Ne~II] line will drop as r$^{-1}$
in the inner region where the density is above the critical
density for the transition and drop as r$^{-3}$ outside the
radius where the density equals the critical density.
This transition would correspond roughly to the position of the measured
source diameter.  For a diameter of 
1.4\arcsec \ and terminal velocity of 17
km s$^{-1}$, the rate at which mass flows through this radius
(the mass-loss rate for the star) is 3$\times$10$^{-5}$ 
M$_{\odot}$ per year.

\subsection{Related IR and Millimeter/Submillimeter Observations}

Because Mon R2 is relatively nearby and has a larger physical size
than most UCHII regions, it has a large angular size.  This extent
makes it possible to resolve thoroughly the structure of the ionized
region and to get some idea of how the ionized gas relates to the
dense core of the surrounding molecular cloud.  We can also investigate
the relationship between the emission nebula and the
continuum sources and scattered continuum emission seen in the 
near-IR.  In this section, we discuss the near-IR continuum and
millimeter/submillimeter line and continuum observations and 
their relationship to our [Ne~II] observations of the ionized gas.
 
The near-IR sources IRS1 and IRS2 both lie within the boundaries of
the nebula 
but IRS1 is the most likely source of ionization.
The highest emission measure regions are close to this source and the
radio continuum and neon brightness along the southeast ridge and the
shell drop with distance from IRS1.
The strength and orientation of the K band polarization vectors in
Mon R2 indicate, however, 
that IRS2 is the main source of scattered light at 2
$\mu$m and that IRS2 also
lies within the nebula. This scattered light dominates
the broad-band continuum in the near-IR 
everywhere except along the southeastern
bright ridge where light from IRS1 makes a significant  contribution
(Hodapp 1987, Aspin and Walther 1990, Yao et al. 1997).
The steeply rising spectrum
of IRS2 (K-L=4.4, Howard et al. 1994) means that much of its
K band continuum may not be photospheric in origin, so that the
dominance of IRS2 at 2 $\mu$m is not a compelling argument for
this source as the primary source of nebular ionization.
 
The bright southeast ridge of ionized gas seen in [Ne~II]
coincides with the northwest boundary of
a dense bar of neutral gas running southwest to northeast (Henning, Chini,
\& Pfau 1992, Gonatas, Palmer \& Novak 1992, Tafalla et al. 1997,
Giannakopoulou et al. 1997).  The dense 
neutral bar is  about
60\arcsec \ long with extensions along the direction of the [Ne~II]
emission ridges to the northwest at its two ends. These extensions
and the bar form the walls off of which the near-IR radiation scatters.
  In the emission lines of molecules with large dipole moments
 like CS, the southeast ridge appears to be sliced through
by a gap immediately to the southeast of IRS1 (Tafalla et al. 1997,
Choi et al. 2000).
 
The massive molecular outflow seen in the low-J CO lines has 
existed for many expansion lifetimes of the ionized nebula.  It has a
crossing time of $\sim$10$^5$ years (Wolf et al. 1990).  The inner
arc minute of this flow includes some very dense material in limb-brightened
shells opening to the northwest and southeast and centered on the bright
bar (Tafalla et al. 1997).

\subsection{Is the Nebula the Ionized Inner Boundary of a Molecular Outflow?}
 
The existing molecular line observations of Mon R2 are not detailed
enough for us to see exactly how the distribution and kinematics
of the ionized and neutral material fit together.
There is some evidence, however, for a relationship
between the inner part of the molecular outflow and the 
ionized gas we see in
[Ne~II].
The velocity extent of the CO emission in the inner
part of the outflow (Meyers-Rice and Lada 1991, Giannakopoulou et al.
1997) and the velocity range over which we see the brightest [Ne~II]
emission are very similar. In addition,
the ionized shell seen at intermediate velocities in the [Ne~II] 
emission nestles inside the northwestern limb-brightened shell
seen by Tafalla et al. (1997) in the CS emission.  

 Tafalla et al. (1997, Figure 15) picture the Mon R2 molecular outflow as a
pair of hyperboloidal or paraboloidal
 streams emerging from an exciting star or
its disk located within the southeastern molecular bar. If IRS1
ionizes the inner edge of this stream, the neon emission will trace
the motions of the material. The velocity patterns seen in the
PV cuts in Figure \ref{pvud}, in particular those through the 
interior of the nebula where the broad line in the southeast
transitions rapidly into a double line whose velocity drops slowly
as one moves north, match well to what one would
expect from one lobe of a  hyperboloid with its axis of symmetry lying almost
in the plane of the sky.  

%
%
 
The picture of the Mon R2 UCHII region as the ionized core of a 
massive molecular outflow has as its particular strength its ability 
to reconcile the short dynamical lifetime of the UCHII region with the
long lifetime of the CO outflow.  One weakness it has, however, is
the lack of symmetry about the likely ionizing source, IRS1.
Unlike the northern outflow lobe, the dense part of the southern lobe
traced by high velocity CS emission (Tafalla et al. 1997) does not appear
to enclose any ionized gas.
Existing observations, however, do not exclude a weaker ionized region
southeast of the molecular bar. The
VLA radio continuum observations do not exclude emission within
the southeast neutral shell
if the surface brightness is any more than a factor of three below
the typical brightness along the northwestern shell. However,
a recent 18\arcsec \ long strip-map in [Ne~II] (K. Allers, private communication)
shows that there is no emission at a level $>$2\% of the peak in
the southeast ridge as far as 12\arcsec \ southeast of the ridgeline.

If there really is no ionized material in the southeast, one
solution to the problem is to blame the outflow on an as yet 
undetected source buried within the southeastern molecular bar . 
A separate, lower luminosity source as the 
origin of the outflow would explain the absence of ionization within
the southern outflow lobe since not much UV 
 radiation from IRS1 
could reach the southern lobe through the channel in the southeastern
bar.
IRS1 would ionize the interior of the outflow's northern lobe.
The enormous
power of outflows around low and intermediate mass YSOs
(Bachiller 1996) makes it unecessary for the highest luminosity source,
IRS1, to generate the flow.  The presence of IRS1, IRS2, and an 
H$_2$O maser source to the northwest (Rodriguez \& Cant\'o 1983)
near the center of the CO flow argue that an additional, more
deeply embedded source in the southeast ridge is not wholly improbable.
 
\section{Conclusions}     

The small thermal width of the [Ne~II] line and the high dynamic
range, sensitivity to both extended and compact spatial components,
 and good velocity resolution of the TEXES observations
permit us to map the distribution and bulk motions of the ionized gas
in Mon R2.  The technique of high spectral resolution [Ne~II] mapping
offers us a way to discover and study in detail the kinematics of
UCHII regions in a way which was not possible with
interferometric observations of radio recombination lines.

Individual spectral lines toward
the ultracompact HII region in Mon R2 
are as narrow as 8 km s$^{-1}$.
Throughout almost all of the source, single or double 
component fits to the spectra imply that turbulence and
small-scale motions
are small compared to the sound speed.
The source exhibits radial motions relative to the quiescent component
of the surrounding neutral medium of $\sim \pm$10 km s$^{-1}$.
A radial expansion model gives an age for the source
of only 4000 years.  As for many other UCHII regions, this 
expansion age is an order of magnitude smaller than the average
UCHII region age implied by the number of such sources and the
OB star formation rate in the Galaxy.

A picture in which the UCHII region represents the ionized
inner rim of a giant molecular outflow fits the morphology and 
kinematics of the region at least qualitatively while, at the same time, 
offering an 
explanation for the large difference between the dynamical timescale
of the ionized region and that of the large-scale CO outflow.
This picture will need to be tested with more detailed observations
of the dense quiescent and outflowing molecular gas near the 
ionized region.

In order to develop a clearer picture of what lies behind the
UCHII region lifetime paradox and more generally of the interaction
of UCHII regions with the dense cores in which they are born, we
have begun to examine a larger sample of such sources.  The sample
will include regions with ionizing stars with a range
of Lyman continuum photon fluxes and with a variety of radio 
continuum morphologies.  We are also observing
other mid-IR fine-structure lines such as the [ArIII] line at
8.99 $\mu$m and the [SIV] line at 10.51 $\mu$m to measure variations
in the hardness of ionizing radiation within UCHII region complexes.

We thank T. Greathouse and the day and night staff at the NASA-IRTF for their 
invaluable assistance.  This work was supported by the National Science
Foundation and by grant
TARP 00365-0473-1999 from the Texas Advanced Research Program.
DJ acknowledges support from the Alexander von Humboldt Stiftung and
the Max-Planck-Institut fuer extraterrestrische Physik during the 
final phases of this project. This research made use of NASA's 
Astrophysics Data System.

\clearpage

\begin{figure}
\epsscale{1.0}
\plottwo{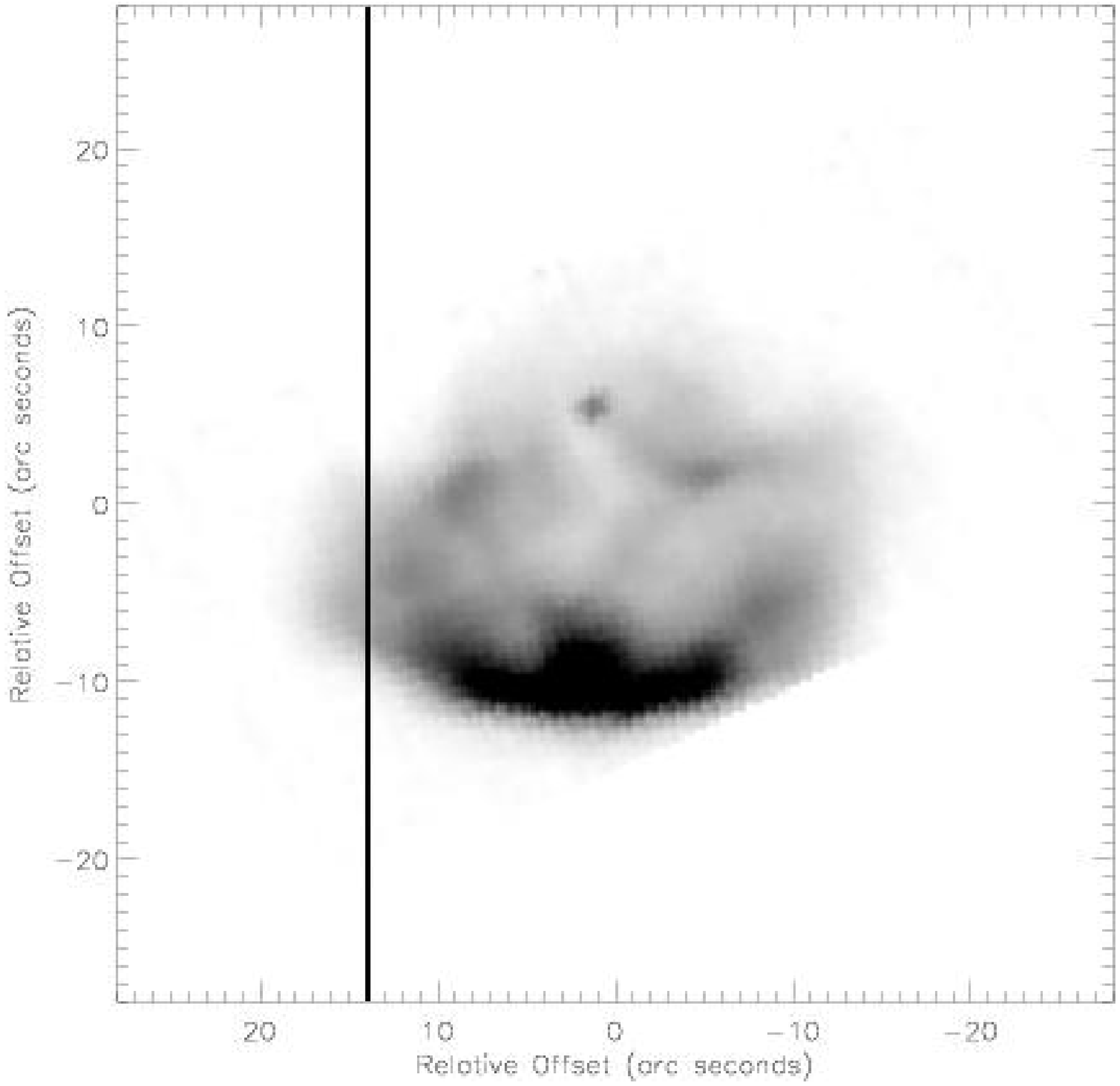}{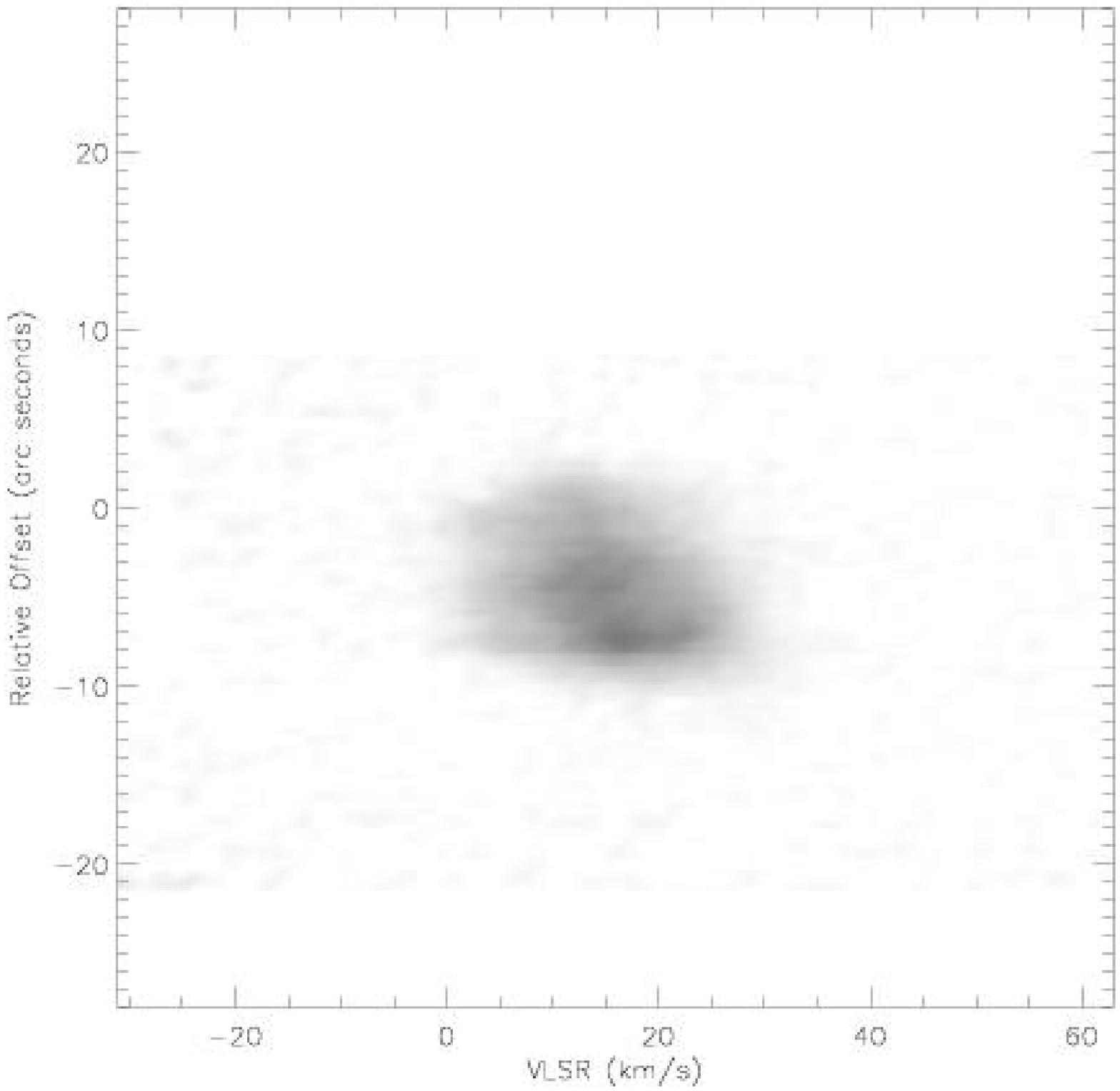}
\end{figure}
\begin{figure}
\epsscale{1.0}
\plottwo{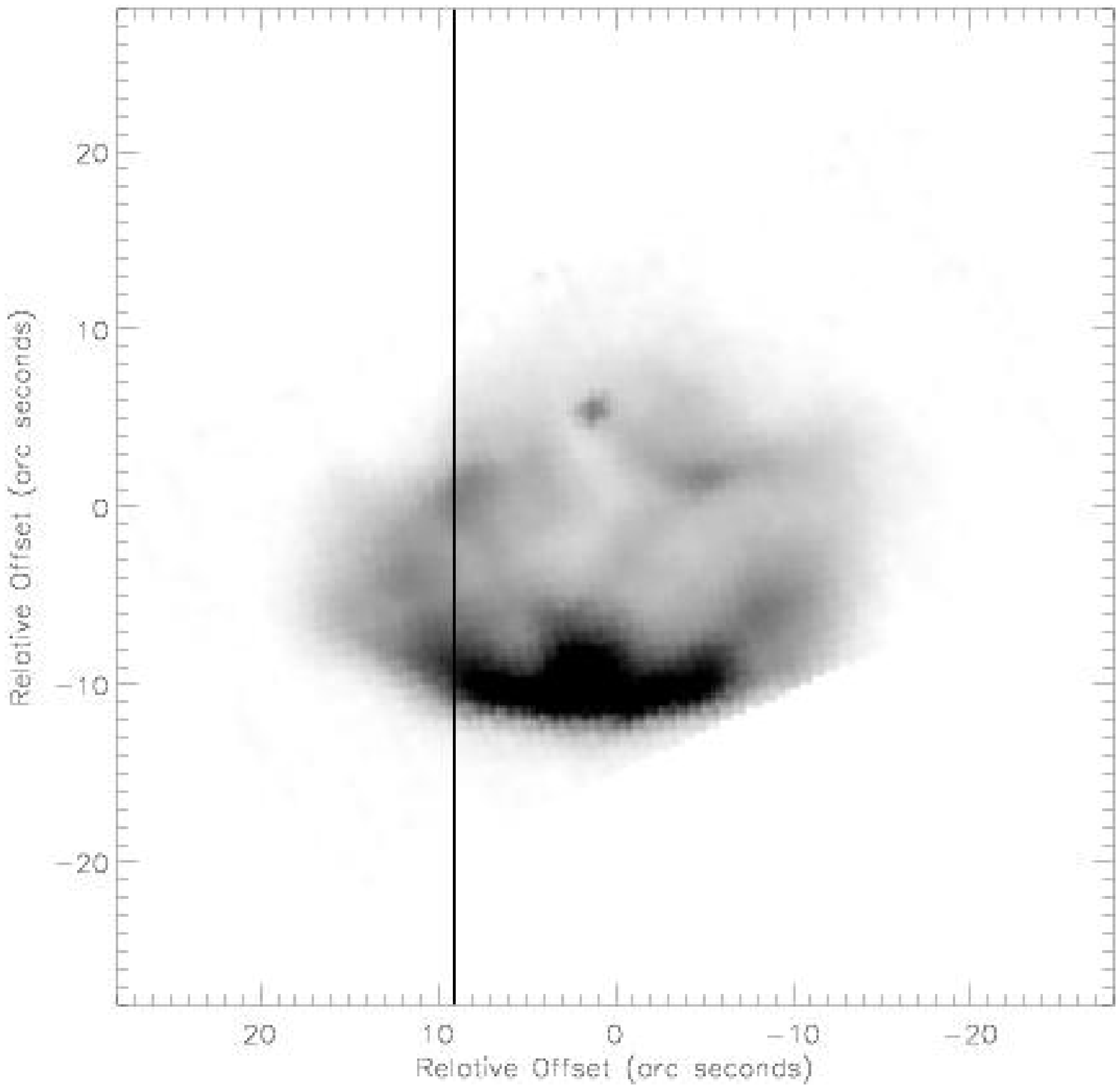}{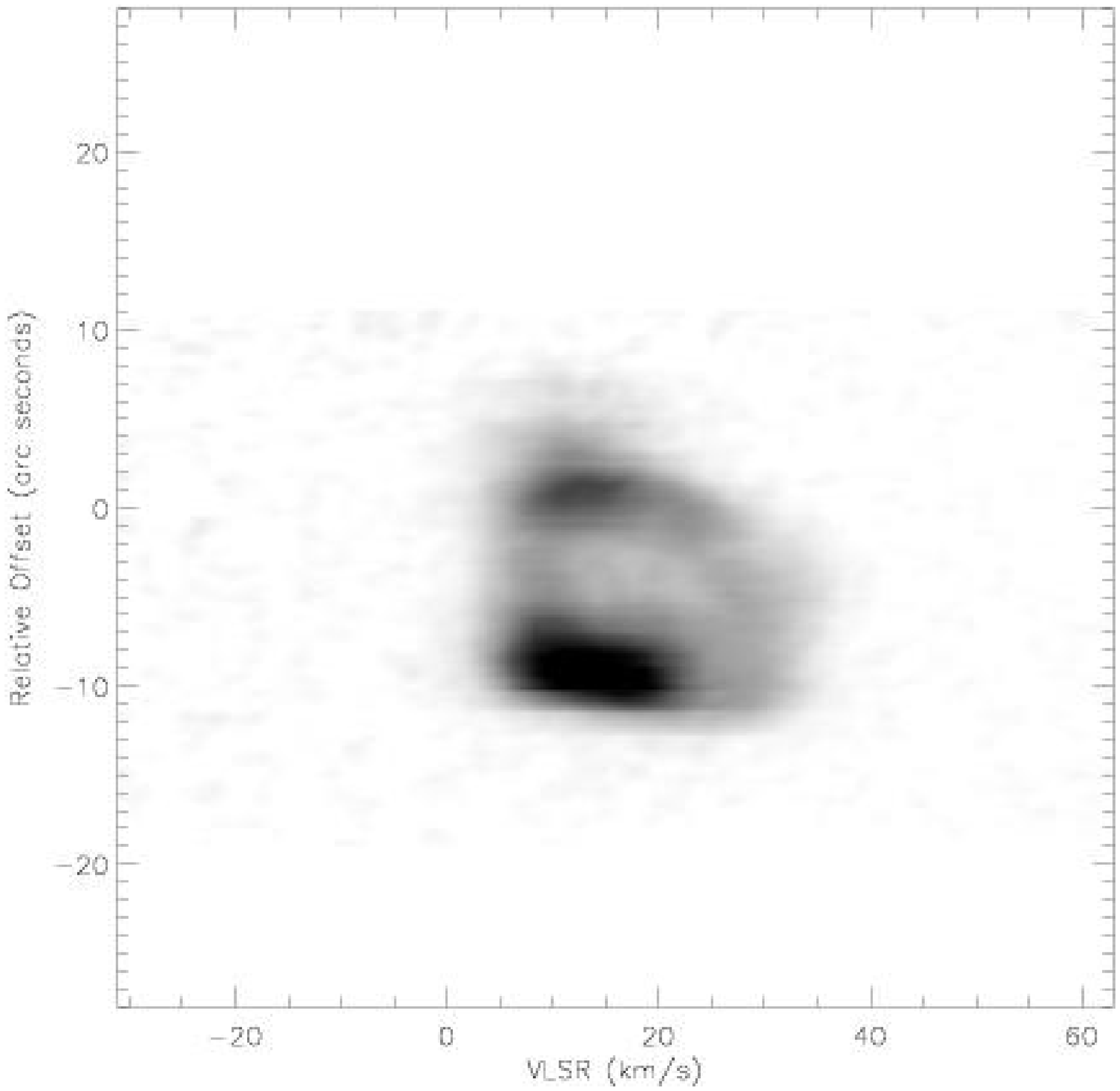}
\caption[3]
{\label{pvud} Position-velocity diagrams for [Ne~II] in Monoceros R2. 
We have rotated the
maps by -25$^{\rm o}$ to make the southeastern ridge horizontal.  In each row, 
the black line in the
left panel shows the location of the PV slice shown on the right.
The integrated intensity scale in the left-hand panels is as in Figure 1.
In the right-hand panels, the scale is linear and white=0 while 
black = 2 erg cm$^{-2}$ sec$^{-1}$ sr$^{-1}$ (cm$^{-1}$)$^{-1}$.
}
\end{figure}
\begin{figure}
\epsscale{1.0}
\plottwo{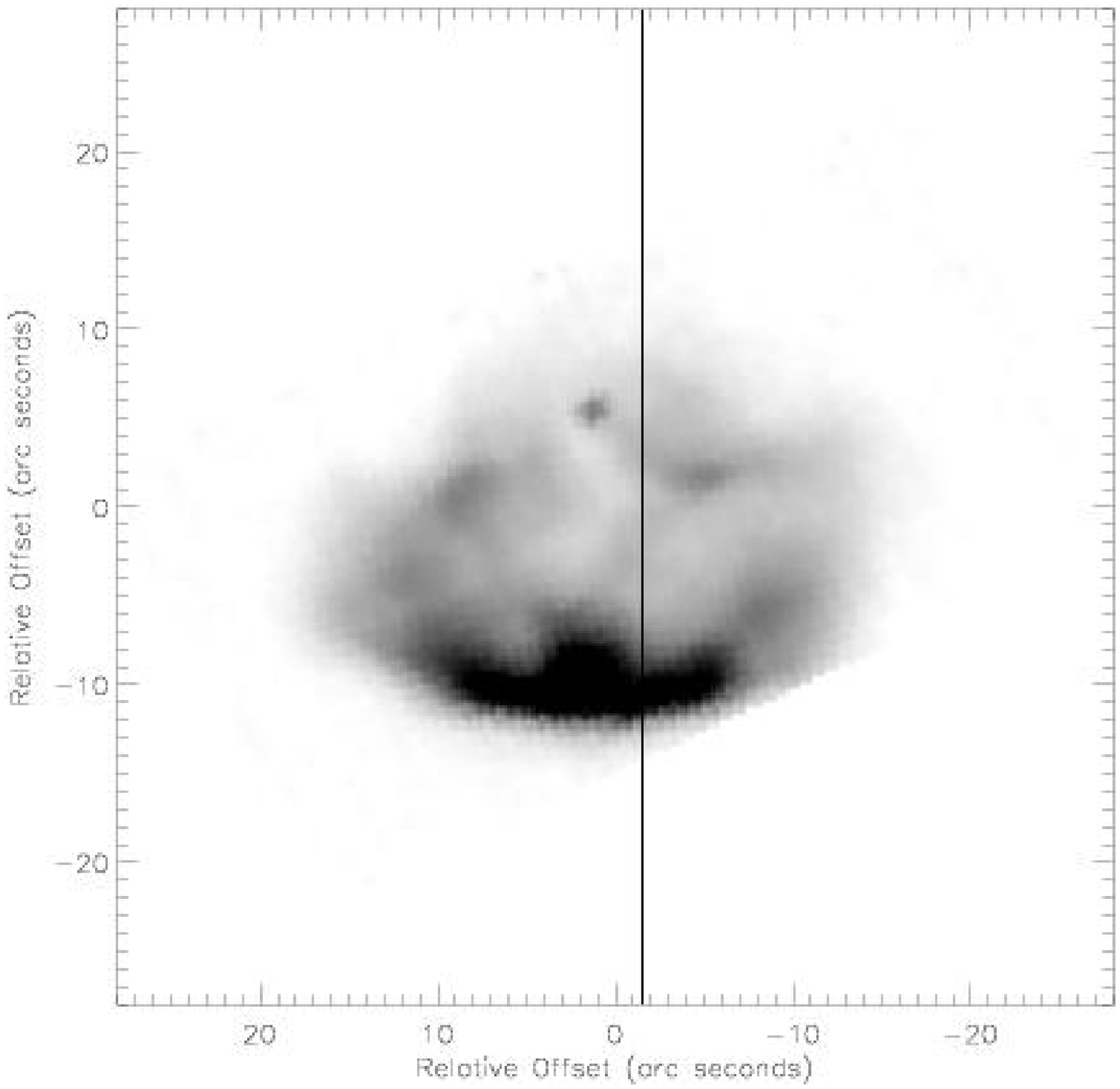}{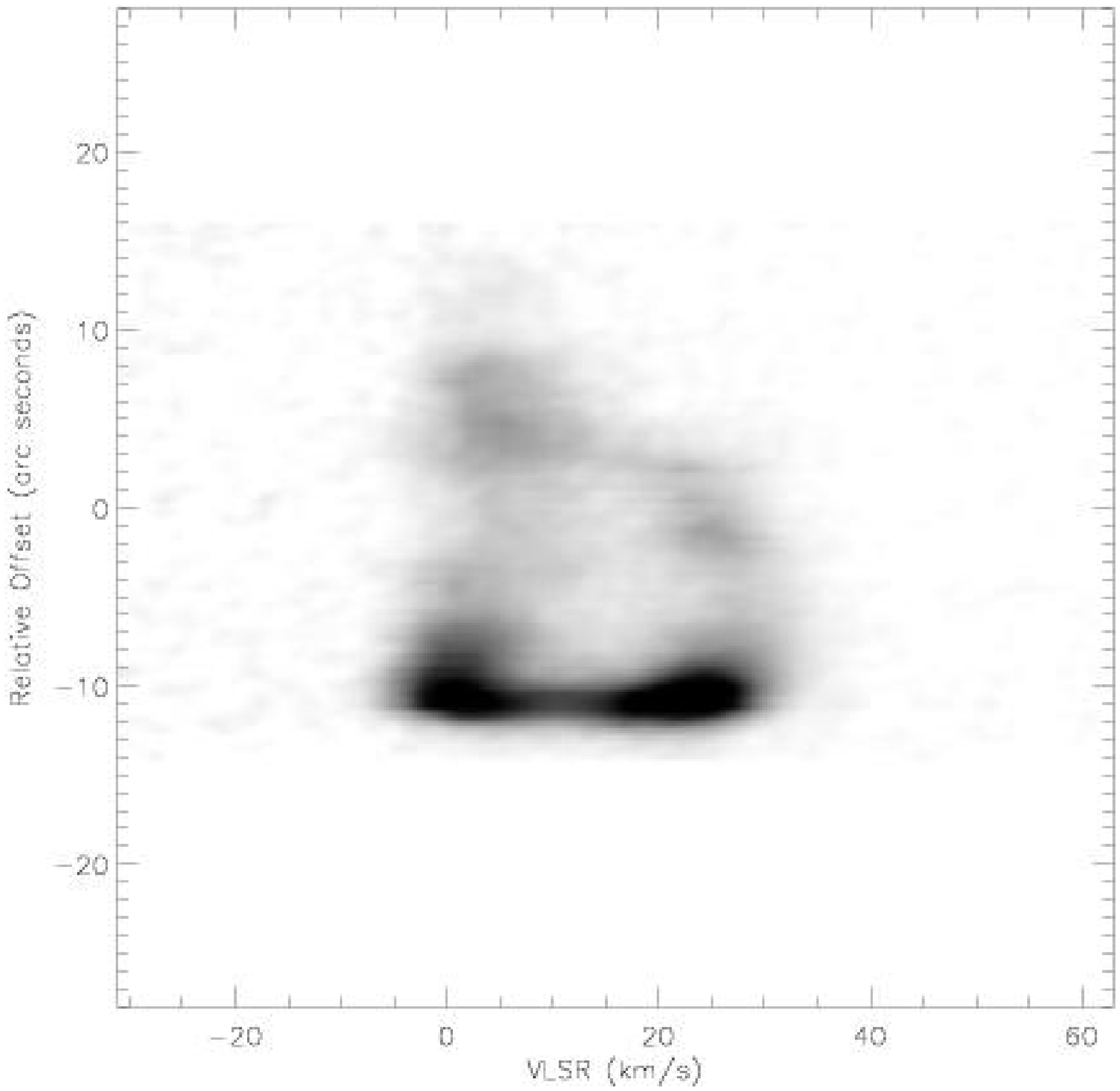}
\end{figure}
\begin{figure}
\epsscale{1.0}
\plottwo{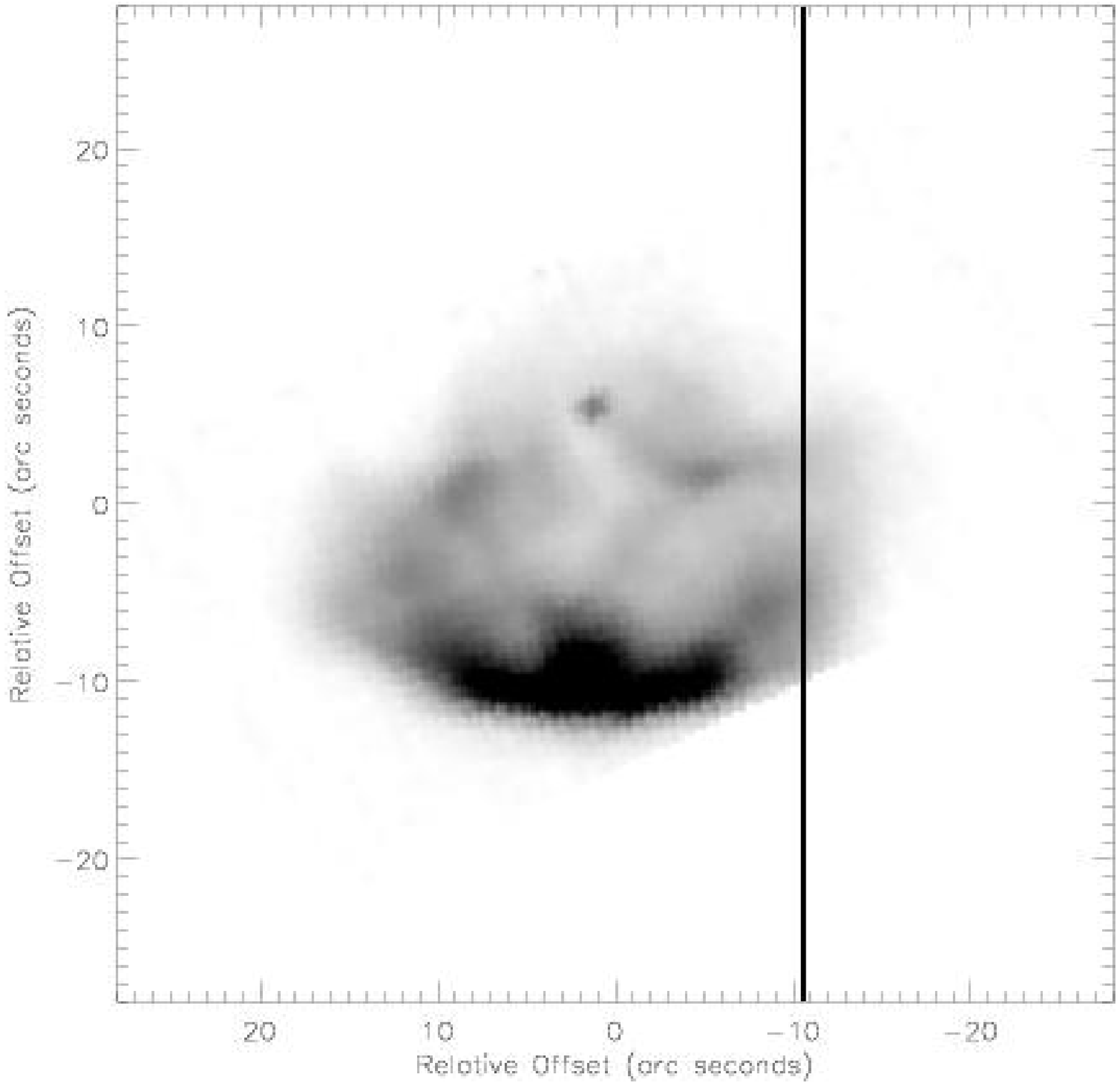}{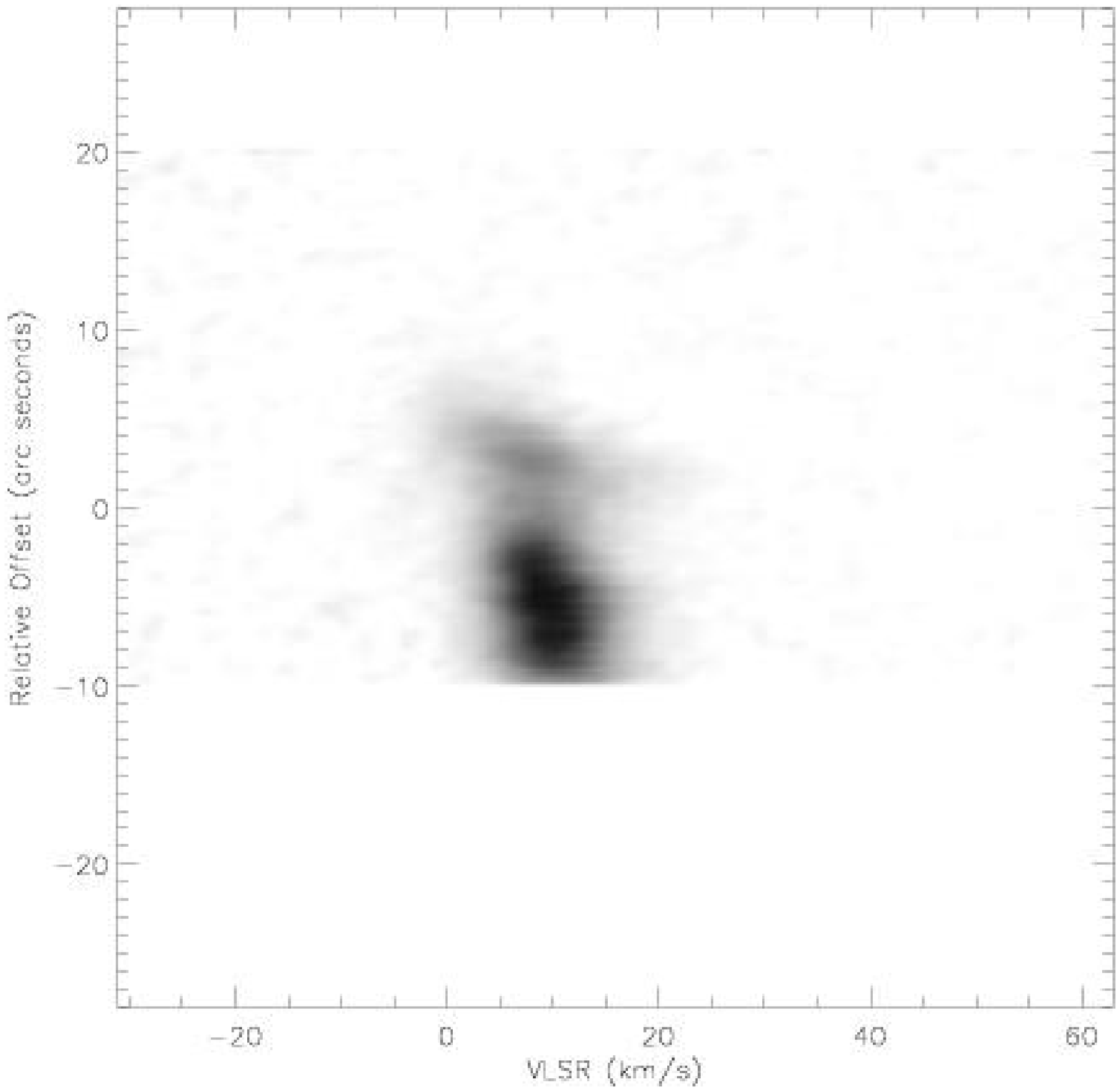}
\caption[3]
{\label{pvud2} Additional position-velocity diagrams for [Ne~II] in Monoceros R2
along cuts perpendicular to the bright southeastern ridge. The intensity scales are as in Figure 5.
}
\end{figure}
\begin{figure}
\epsscale{1.0}
\plottwo{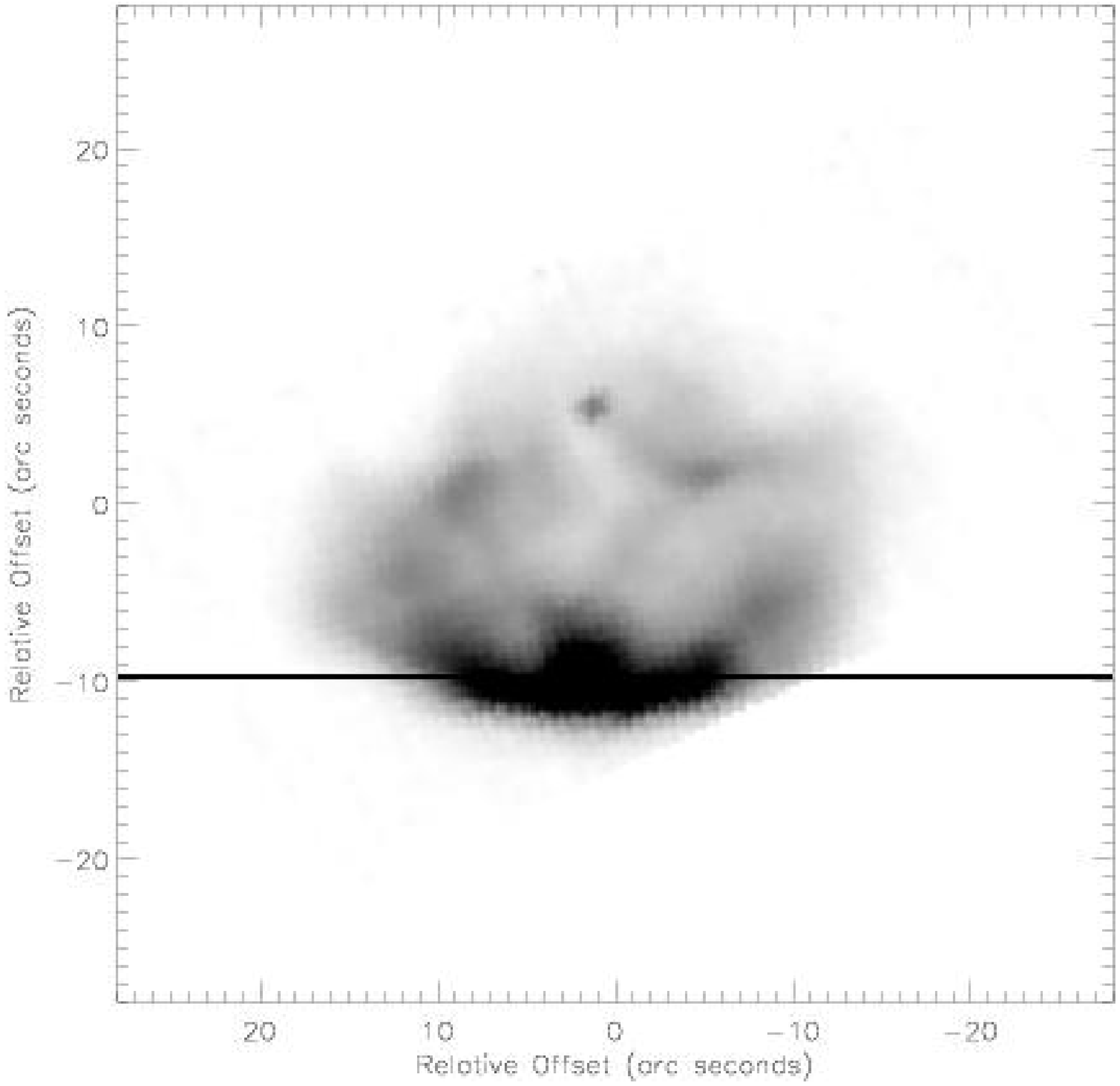}{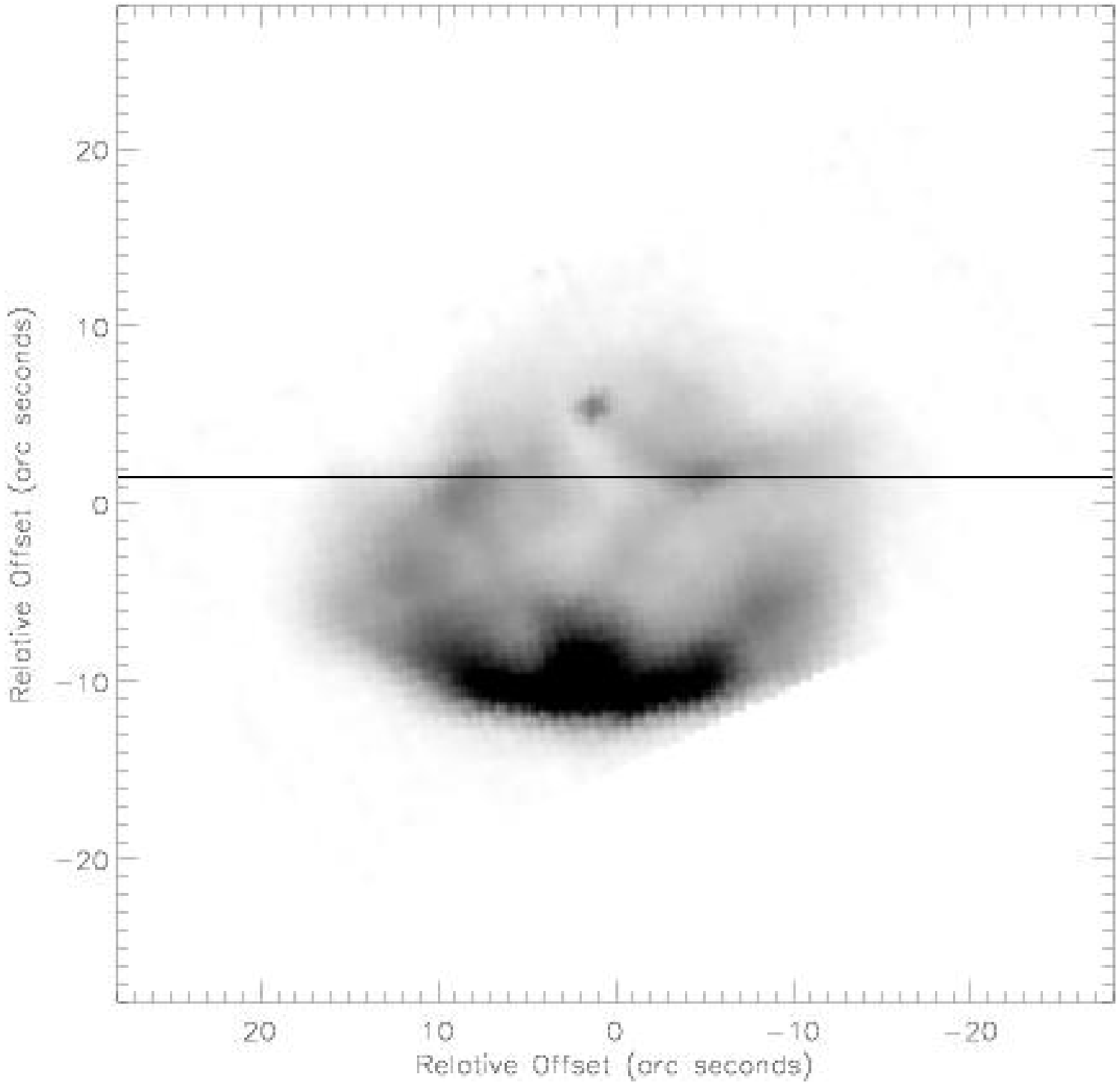}
\end{figure}
\begin{figure}
\epsscale{1.0}
\plottwo{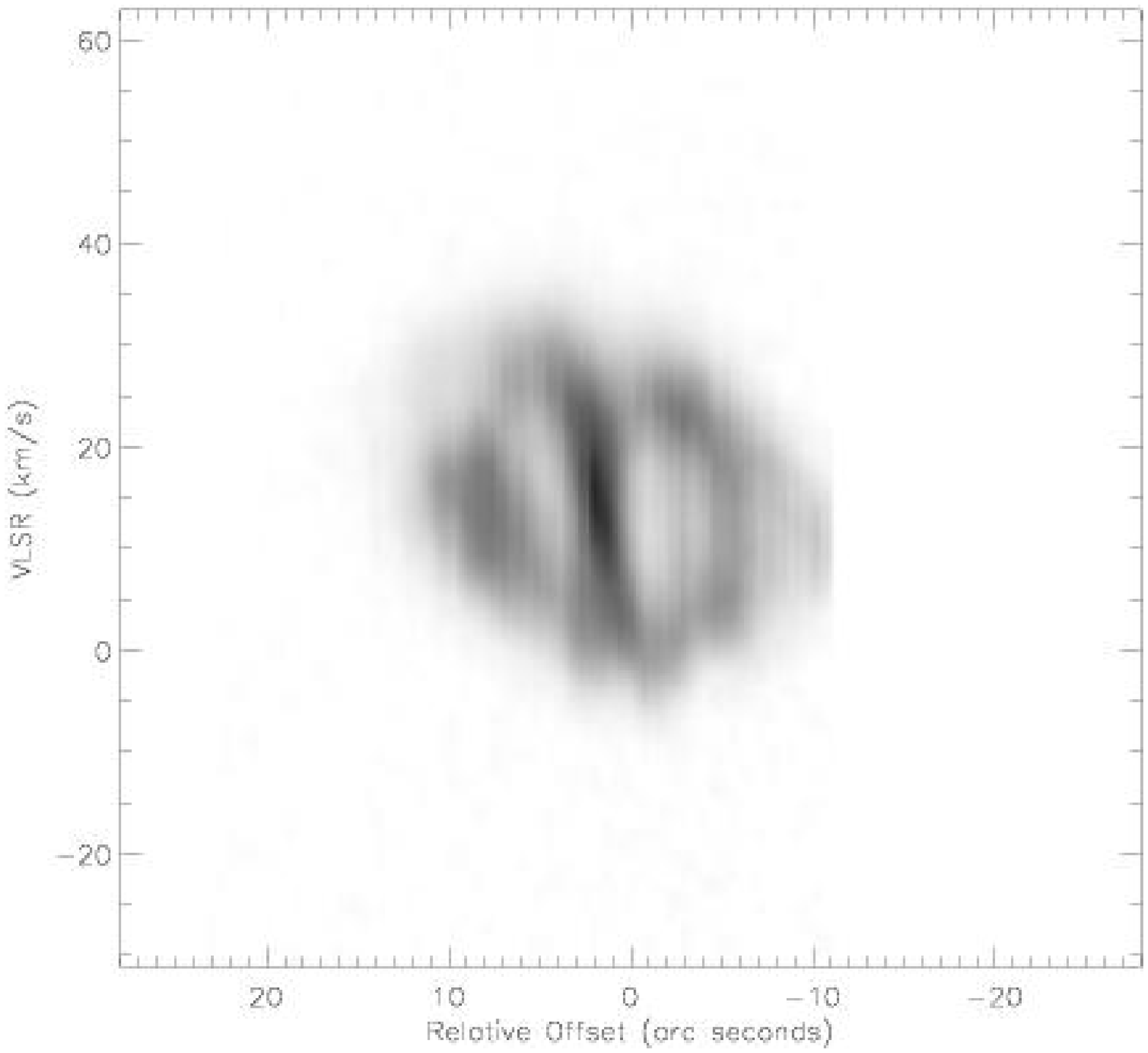}{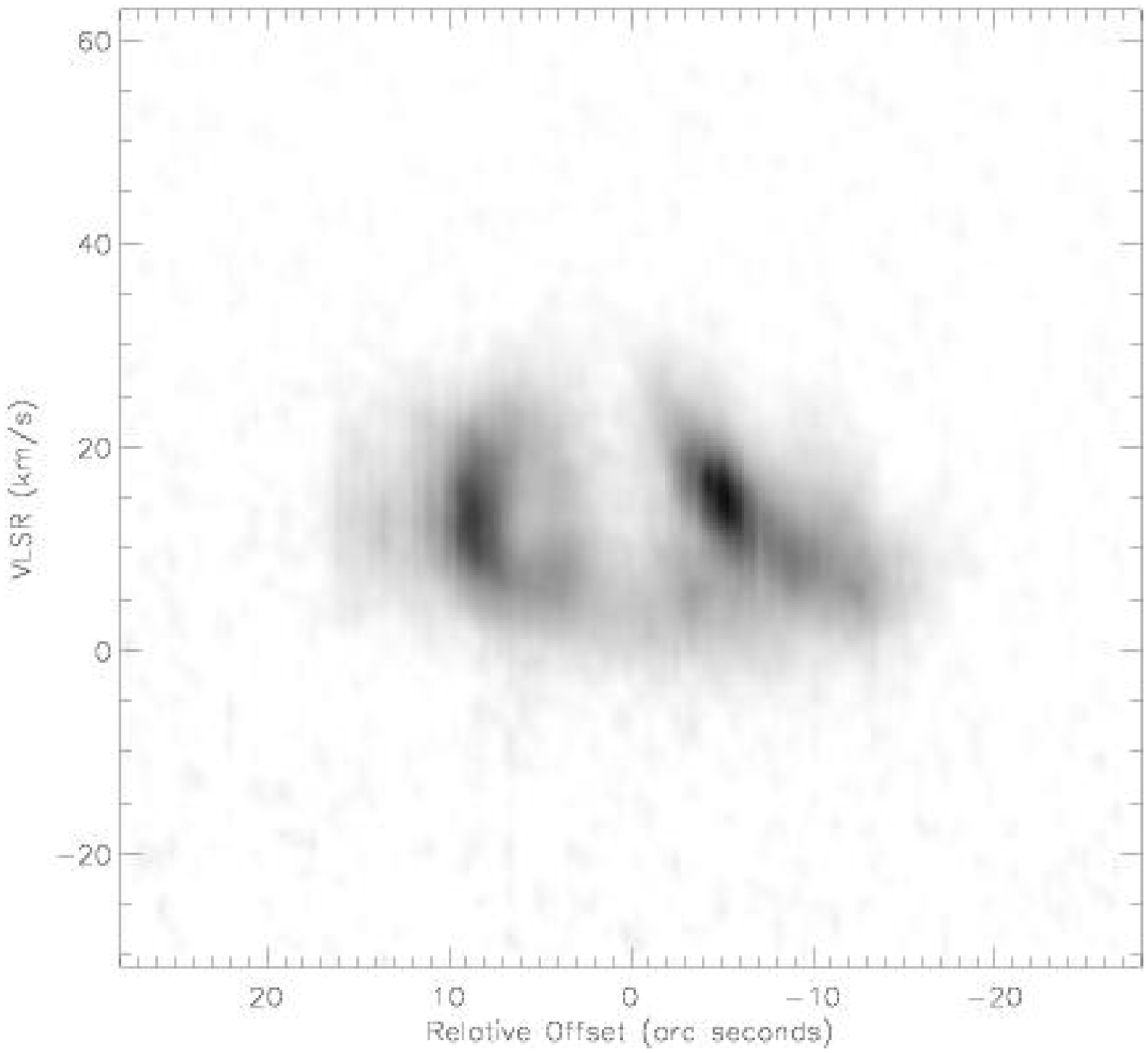}
\caption[3]
{\label{pvlr} Position-velocity diagrams along cuts parallel to the 
southeastern ridge. For the lower right panel, the intensity scales 
are as in Figure 5. In each column, the black line in the upper panel shows
the location of the PV slice shown in the lower panel.
For the lower left panel, black = 5 erg cm$^{-2}$ sec$^{-1}$ sr$^{-1}$ (cm${-1}$)$^{-1}$.
}
\end{figure}

\clearpage

\end{document}